# Molecular Dynamics Simulations of Carbon Nanotubes as Gigahertz Oscillators

S. B. Legoas[1], V. R. Coluci[1], S. F. Braga[1], P. Z. Coura[2], Sócrates O. Dantas[2], and D. S. Galvão[1]

[1]*Instituto de Física "Gleb Wataghin", Universidade Estadual de Campinas, 13083-970 Campinas, SP, Brazil and*
[2]*Departamento de Física, ICE, Universidade Federal de Juiz de Fora, 36036-330 Juiz de Fora, MG, Brazil*

(Dated: September 3, 2002)

Recently Zheng and Jiang [PRL **88**, 045503 (2002)], based on static models, have proposed that multiwalled carbon nanotubes could be the basis for a new generation of nanooscillators in the several gigahertz range. In this work we present the first molecular dynamics simulation for these systems. Different nanotube types were considered in order to verify the reliability of such devices as gigahertz oscillators. Our results show that these nanooscillators are dynamically stables when the radii difference values between inner and outer tubes are of ∼ 3.4 Å. Frequencies as large as 38 GHz were observed, and the calculated force values are in good agreement with recent experimental investigations. Moreover, our results contradict some predictions made by Zheng and Jiang.

The technological advances have led to the need of creating functional devices at nanometer scale [1]. There have been continuing efforts in fabricating nanomechanical systems operating in high frequencies but gigahertz range is beyond our present micromachining technology [2, 3].

A breakthrough in this area has been obtained by Cumings and Zettl [4]. They demonstrated the controlled and reversible telescopic extension of multi walled carbon nanotubes (MWNT), thus realizing ultralow-friction nanoscale linear bearing (Fig. 1a). They also demonstrated that repeated extension and retraction of telescopic nanotube segments revealed no wear or fatigue on the atomic scale [4]. This type of structure was foreseen by E. Drexler [1, 3] and remained a *Gedanken* experiment until Cumings and Zettl experimental realization.

Depending on the distance separation between two adjacent nanotubes it is possible to have an almost perfect sliding surface [5]. The van der Waals interactions between the nanotubes create a restoring force that cause the inner tube to retract (Figs. 1 and 2) and can be the physical basis (and within our present technological capabilities) to build nanodevices such as nanobearings, nanosprings, and nanoswitches [1, 2, 4, 6, 7].

Using static models on a slightly modified Cumings and Zettl set up (Fig. 1b), Zheng and collaborators [2, 8] have shown that multishell nanotubes could lead to nanooscillators in the range of several gigahertz. However, in order to demonstrate that these devices can be fully functional, dynamical aspects such as temperature and force and energy temporal fluctuations (once they can compromise device performance) have to be considered.

In this work we present the first molecular dynamics study for these systems. Our results show that telescopic extension and retraction movements are possible for a large combination of tube diameters and types (armchair, zig-zag, chiral, and combinations). However, sustained oscillations (necessary condition for oscillator devices) are only possible when the radii difference values between inner and outer tubes (Fig. 2) are of ∼3.4 Å, independently of tube type. We have observed that these oscillations (in gigahertz scale) are possible for a large range of temperature (up to 400 K), which strongly suggest that in fact multishell nanotubes could be the basis for a new generation of nanooscillators. Also, some predictions made from static models [2, 8] are contradicted by our dynamical model, thus demonstrating the importance of taking into account temperature effects for these devices.

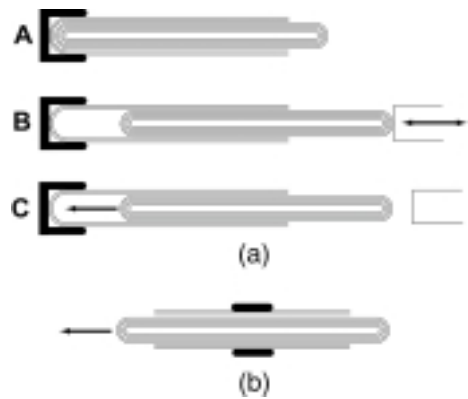

FIG. 1: (a) Schematic representation of Cumings and Zettl experimental set up [4] using a high-resolution transmission electron microscope. The A, B, and C sequences show the experimentally observed controlled core nanotube reversible telescoping. (b) Schematic set up of the multiwalled carbon nanotubes oscillators proposed by Zheng and coll. [2, 8] See text for discussions.

We have carried out molecular dynamics simulations in the framework of classical mechanics with a standard molecular force field [9] which includes van der Waals, bond stretch, bond angle bend, and torsional rotation terms. We have considered structures containing up to 6,000 carbon atoms which precludes the use of quantum methods. This methodology has been proven to be very effective in the study of dynamical properties of carbon structures [10].

For all simulations the following convergence criteria were used: maximum force of 0.005 kcal/mol/Å, root mean square (RMS) deviations of 0.001 kcal/mol/Å, energy differences of 0.0001 kcal/mol, maximum atomic displacement of 0.00005 Å, and RMS displacement of 0.00001 Å. After initial minimization procedures selective microcanonical (constant number of particles, volume and total energy) impulse dynamics was used. Time steps of 1 femto seconds were used for all simulations.

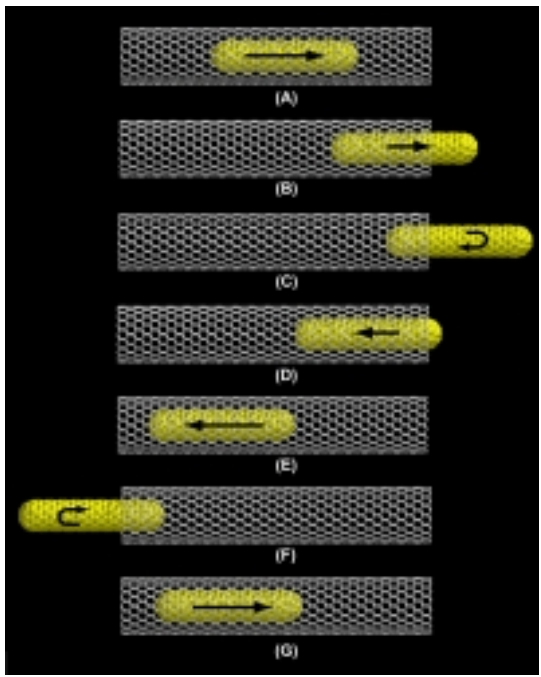

FIG. 2: Snapshots from the molecular dynamics simulations showing the oscillatory behavior of a (9,0) nanotube inside a (18,0) one with its double length. The associated force and potential energy profiles are presented in Figs. 3a and 3b. See text for discussions.

The structures were generated in the following way: a single wall nanotube (SWNT) of lenght $L$ closed at both ends acting as a moveable core and a SWNT or MWNT of lenght $L'$ opened at their ends (one or both) multi-shelling the closed SWNT (Fig. 2). The generated nanotubes are isolately optimized (geometries and energies) and reoptimized when shelled. After the reoptimization process, the impulse dynamics is applied attributing an initial velocity to the internal closed SWNT in order to initialize the movement (Fig. 2). We have analyzed inner and outer arm-chair, zig-zag, chiral tubes and their combinations of different lenghts and initial impulses.

Our results showed that external SWNT or MWNT produce the same dynamical general features. The net effect of adding extra shells to the external MWNT is to make the more internal shells stiffer. This effect saturates in terms of oscillatory frequency with three external nanotubes. Thus, in order to save computational effort and to better illustrate some dynamical aspects, as well as to provide a more direct comparison with the results from static models [2, 8], we will only present and discuss below the results for the cases of two shelled nanotubes, where the external one is kept fixed and the internal is freely to move in all directions.

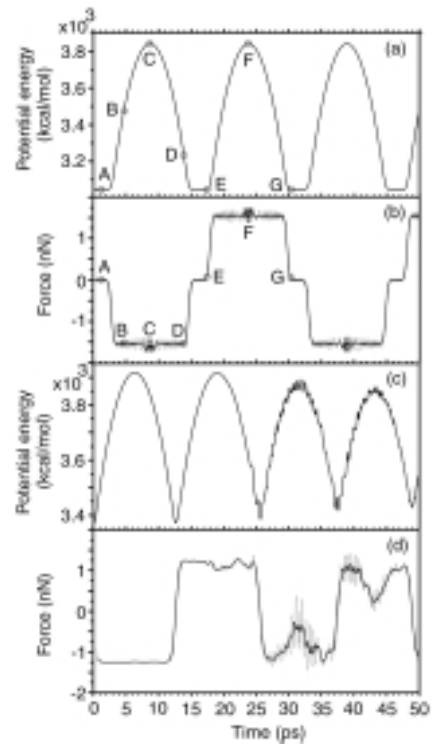

FIG. 3: Force and Potential Energy results from molecular dynamics simulations. (a) and (b) refer to the case of a perfect oscillatory behavior (Fig. 2). (c) and (d) illustrate cases where dissipative effects, that compromise the sustained oscillations, are present. A small force fluctuation due to vibrational motion can be observed. A thick line was sketched to represent the mean forces.

The limit value for the initial tube impulse (inner tube ejection) can be estimated using the simple model of considering the kinetic energy described by the motion of the center of mass. The van der Waals force is then given by:

$$F = -\sqrt{\frac{m}{2K}} \frac{dU}{dt} \ , \qquad (1)$$

where $m$ is the mass of the inner tube, and K and U refer to the kinetic and potential energy that it is not explicitly time dependant, respectively.

The maximum mean force, acting on inner nanotube, is given by:

$$F_{max} = \left(\frac{m}{2K_{max} - \dot{U}_{max}\Delta t}\right)^{1/2} \dot{U}_{max} \ . \qquad (2)$$

In Eq. (2), $\dot{U}_{max}$ is the maximum value of the time derivative of potential energy $U$, $K_{max}$ is the maximum mean kinetic energy (when the excess van der Waals interaction is zero), and $\Delta t$ the elapsed time for van der Waals force to reach its maximum (from zero value). In this way we can estimate the oscillating frequency through the following equation:

$$f = \frac{1}{2}\left(\frac{2mv_0}{F_{max}} + \frac{\Delta L}{v_0}\right)^{-1}, \qquad (3)$$

where $v_0$ is the initial core velocity and $\Delta L = L' - L$ is the lenght difference between outer and inner tubes, respectively.

Finally, the limit value for the initial core velocity can be estimated using the expression:

$$v_s = \left(\frac{2LF_{max}}{m}\right)^{1/2}, \qquad (4)$$

where $L$ is the inner tube lenght. For the case of the oscillator in Fig. 2, we estimated the limit initial velocity as being equal to $\sim 1,370$ m/s.

In Fig. 2 we show typical results for cases where the sustained oscillatory motion occurs. Figs. 3a and 3b show the corresponding potential energy and forces. Initially the internal tube has velocity of 1,200 m/s along the tube's axis (Fig. 2A). While the internal tube is shelled by the external one no force is present (Figs. 2A, 2E, and 2G; Fig. 3b, point A). As the internal tube is telescoped the excess van der Waals interaction energy creates a restoring force that opposes the tube extrusion (Fig. 2B). The internal tube has its velocity reduced until complete stop (Fig. 2C). Then its movement is reversed and the force continues acting now helping the intrusion (Fig. 2D) until the core tube is completely shelled again (Fig. 2E, Fig. 3b). The situation is repeated on the opposite side (Fig. 2F) until whole oscillatory cicle is completed (represented in Fig. 2G). See complementary material, video01. The calculated force values are compatible with the ones estimated by Cumings and Zettl [4] and Zheng et coll. [2, 8].

This sustained oscillatory behavior is possible (for all kind of tube combinations) only when a perfect coupling between the tubes occurs (Fig. 4a). For other couplings (Fig. 4b), although the telescopic and retractive movements are possible, dissipative forces (Figs. 3c and 3d) and momentum exchange between the tubes compromise the sustained oscillatory behavior. See complementary material, video02.

The requirement for a perfect coupling between the nanotubes is satisfied for a huge number of possible cases. For external radii from 7 to 200 Å the number of possible configurations is up to $9.5 \times 10^6$. The major contributions ($\sim 98.4$ %) come from combinations involving only chiral tubes of distinct chiralities (Fig. 5).

Zheng et collaborators [8] speculated that closing one end of the external nanotube would increase the oscillatory frequencies as a result of the huge repulsive force over the core nanotube. However our results demonstrated that this would not occur. Due to the interactions between the tubes these close contacts would induce massive transfer momentum between them severely affecting the oscillatory behavior. The system would exhibit a behavior similar to that shown in Figs. 3c and 3d, and no sustained oscillatory regime is possible. See complementary material, video03.

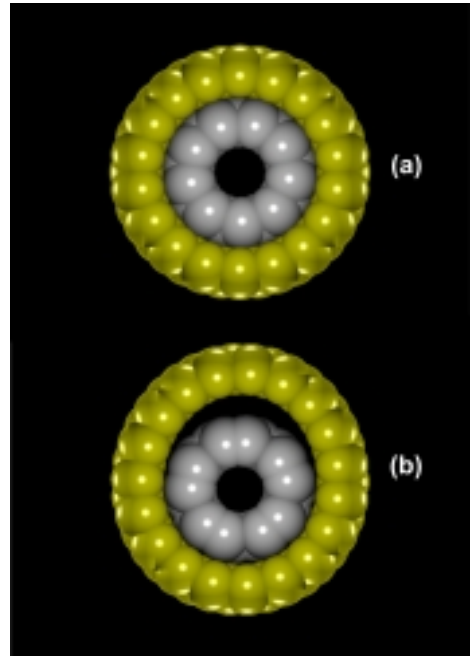

FIG. 4: Cross-section view of the nanotube configurations. (a) (9,0) and (18,0) nanotubes with a perfect fitting allowing a sustained oscillatory movement. (b) (5,5) and (19,0) nanotubes, an imperfect fitting is clearly observed and this prevents the sustained oscillatory regime due to transfer momentum between the tubes (Fig. 3c and 3d).

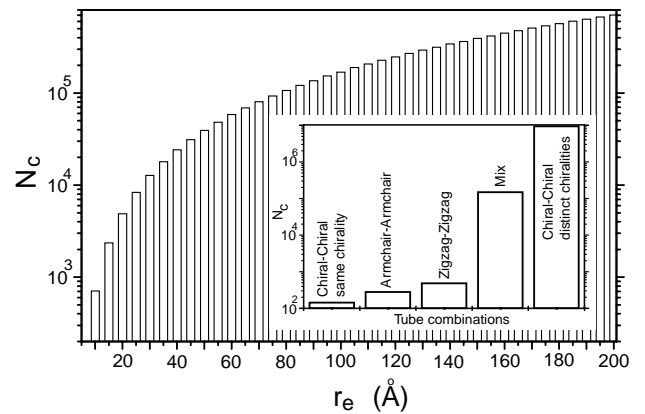

FIG. 5: Number of possible configurations as a function of the external nanotube radius where the radii difference between inner and outer nanotubes is in the range of 3.35 to 3.6 Å. In the inset the number of possible configurations as a function of the tube types is shown.

Our results have shown that multishell nanotubes can be used to make functional nanooscillators in the gigahertz range. However a fundamental question is how to initialize these devices in a controllable way. It has been proposed that this could be achieved applying external electric fields or through charge injection, although it has been recognized that these represent serious technological challenges [1, 2, 4].

Perhaps, a simpler alternative way would be applying variable magnetic fields. The magnetic force (produced by the variable magnetic flux) can be used to initialize the movement in a controllable way. Consider a combination of an internal metallic (or filled with metallic materials) nanotube and an external semiconducting one. In this case the magnetic force will act selectively on the inner nanotube, thus being able to initialize the oscillatory regime [11]. An adapted experimental Aharonov-Bohm set up [12] could be one possibility to realize this approach, specially for a combination of small diameter nanotubes (large diameter nanotubes tend to be metallic).

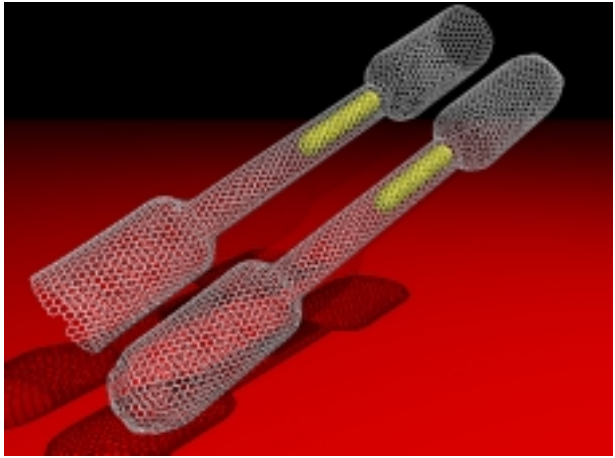

FIG. 6: Schematic artistic representation of the proposed sealed nanooscillator. The core (inner) nanotube oscillates inside a bi-nanotube junction.

One unique aspect of the concept of these nanooscillators is that their operating frequencies can be modulated by the lenght difference between the nanotubes and by the strength of the external excitatory agent. When the external tube is longer than the internal one the frequency of the movement will be determined by two components, one depending on the van der Waals interactions and the other by the required time (when no force acts upon it) to the internal tube to cross the external one (Fig. 3). This energy can be externally modulated and the oscillator will have a variable range of operating frequencies.

In principle the nanooscillators devices can work with the external nanotube having both ends opened. Cumings and Zettl [4] observed that although sometimes a residual amorphous carbon was present inside outer nanotube, such contamination does not seem to significantly affect the bearing action since it is brushed away upon tube reinsertion. However, in terms of working devices, this contamination can create serious problems and sealed structures would be ideal. As the external tube can not have its ends closed (the excess of the van der Waals interactions is the physical basis for the oscillatory regime) one solution is 'to fuse' tubes of different diameters in order 'to seal' the device (Fig. 6). It is important to notice that radii difference values between the joined tubes should be at least of $3 \times 3.4$ Å in order to prevent any undesirable interactions between them. The sealed devices constitute the ideal configuration and work perfectly. See complementary material, video04. The recent advances in the synthesis (SWNT as long as 20 cm [13]), controllable manipulation and chemical etching [14] (designed molecular junctions [15]) of carbon nanotubes make the sealed oscillators we are proposing feasible in our present technological capabilities. We hope the present work could stimulate further studies along these lines.

The authors acknowledge financial support from the Brazilian Agencies CNPq, FAPEMIG, and FAPESP, and the use of computational facilities at CENAPAD/SP.